\documentclass[twocolumn]{aastex63}
\usepackage{amsmath}
\usepackage{here}
\usepackage{graphics}

\newcommand{\rhog}{\rho_{\mathrm{g}}}
\newcommand{\rhod}{\rho_{\mathrm{d}}}

\newcommand{\gas}{\mathrm{g}}
\newcommand{\dst}{\mathrm{d}}

\newcommand{\sigmag}{\Sigma_{\gas}}

\newcommand{\sigmad}{\Sigma_{\dst}}

\newcommand{\tstop}{t_{\mathrm{stop}}}

\newcommand{\taus}{\mathrm{St}}

\newcommand{\cs}{c_{\mathrm{s}}}

\newcommand{\hd}{H_{\mathrm{d}}}
\newcommand{\vk}{v_{\mathrm{K}}}

\newcommand{\vfrag}{v_{\mathrm{frag}}}

\newcommand{\rhoint}{\rho_{\mathrm{int}}}
\newcommand{\tdri}{t_{\mathrm{drift}}}
\newcommand{\vdri}{v_{\mathrm{drift}}}

\newcommand{\tcoag}{t_{\mathrm{coag}}}
\newcommand{\taucoag}{\tau_{\mathrm{coag}}}

\newcommand{\taucoageff}{\tau_{\mathrm{coag, eff}}}
\newcommand{\tSI}{t_{\mathrm{SI,sat}}}
\newcommand{\tauSI}{\tau_{\mathrm{SI,sat}}}

\received{--}
\accepted{--}

\shorttitle{Dust growth during streaming instability}
\shortauthors{Tominaga and Tanaka}

\begin{document}

\title{Rapid dust growth during hydrodynamic clumping due to streaming instability}

\correspondingauthor{Ryosuke T. Tominaga}
\email{ryosuke.tominaga@riken.jp}
\author[0000-0002-8596-3505]{Ryosuke T. Tominaga}
\affiliation{RIKEN Cluster for Pioneering Research, 2-1 Hirosawa, Wako, Saitama 351-0198, Japan}

\author[0000-0001-9659-658X]{Hidekazu Tanaka}
\affiliation{Astronomical Institute, Graduate School of Science, Tohoku University, 6-3, Aramaki, Aoba-ku, Sendai 980-8578, Japan}

%
%
%



\begin{abstract}
Streaming instability is considered to be one of the dominant processes to promote planetesimal formation by gravitational collapse of dust clumps. The development of streaming instability is expected to form dust clumps in which the local dust density is strongly enhanced and even greater than the Roche density. The resulting clumps can collapse to form planetesimals. Recent simulations conducted long-term simulations and showed that such strong clumping occurs in a wider parameter space than previously expected. However, the indicated timescale for strong clumping can be on the order of tens to hundreds Keplerian periods. In this paper, we estimate the growth time of dust grains during the pre-clumping phase. We find that the dust growth considerably proceeds before the strong clumping because even the moderate clumping due to streaming instability increases the local dust-to-gas ratio $\gtrsim10$. Depending on the gas sound speed, the dust collision velocity can be kept below $\sim 1\;\mathrm{m/s}$ once sufficiently strong dust clumping occurs. Thus, even silicate grains might have the potential to grow safely toward the size whose Stokes number is unity during the clumping. Our results demonstrate the importance of local dust coagulation during the dust clumping due to streaming instability. 
\end{abstract}



\section{Introduction}\label{sec:intro}

There are two promising processes for planetesimal formation: direct collisional growth of dust grains \citep[e.g.,][]{Ormel2007a,Okuzumi2012,Kataoka2013b,Krijt2016,Arakawa2016,Garcia2020,Kobayashi2021} and gravitational collapse of a dust layer \citep[e.g.,][]{Safronov1972,Goldreich1973,Sekiya1983,Youdin2002}. The latter process is expected to occur once a sufficient amount of dust is concentrated against turbulent diffusion in a disk \citep[e.g.,][]{Cuzzi1993,Sekiya1998}. Some dust-driven instabilities have been proposed as the dust concentration mechanism \citep[e.g.,][]{YoudinGoodman2005,Johansen2007,Youdin2011,Takahashi2014,Tominaga2021,Tominaga2022a,Tominaga2022b}. Among the dust-driven instabilities, streaming instability has been extensively studied analytically and numerically \citep[e.g.,][]{YoudinGoodman2005,Youdin2007,Johansen2007,Johansen2007nature,Krapp2019,Chen2020,Umurhan2020,Paardekooper2020,Paardekooper2021,McNally2021,Zhu2021,Yang2021,Carrera2021,Carrera2022}. Streaming instability occurs on a spatial scale much smaller than the gas scale height. The nonlinear evolution causes turbulent motion and transient/intermittent formation of azimuthally elongated filamentary structures \citep[e.g.,][]{Johansen2007}. The resulting dust-dense regions collapse self-gravitationally once the local dust density exceeds the Roche density $\rho_{\mathrm{R}}$ \citep[e.g., ][]{Johansen2007nature,Simon2016} and the self-gravity overcomes turbulent diffusion \citep[][]{Gerbig2020,Klahr2020,Klahr2021}. This combined process of streaming instability and the subsequent self-gravitational collapse is expected to explain planetesimal formation \citep[e.g.,][]{Drazkowska2016,Nesvorny2019,Gole2020,Gerbig2020}.

 Previous studies investigated the condition for streaming instability to cause strong dust clumping where  the maximum dust density $\rho_{\dst,\mathrm{max}}$ exceeds the Roche density (e.g., $\sim$ several hundred times gas density for low-mass disks)\footnote{ The Roche density is $\rho_{\mathrm{R}}\equiv9\Omega^2/4\pi G\simeq 3\times10^2\rhog\times(Q/55)$, where $\Omega$ is the Keplerian angular velocity, $G$ is the gravitational constant, $\rhog$ is the midplane gas density, and $Q$ is the Toomre's $Q$ value of the gas disk.  }, which is necessary for the self-gravitational collapse of the dust clump \citep[][]{Carrera2015,Yang2017,Li2021}. \citet{Carrera2015} and \citet{Yang2017} found that the dust-to-gas surface density ratio should be a few times higher than 0.01 for the strong clumping, which also depends on the Stokes number of dust, $\taus$. \citet{Gerbig2020} discussed that the critical value for the gravitational collapse to proceed also depends on the gas pressure gradient and the Toomre's $Q$ value of the gas disk. Recently, \citet{Li2021} revisited the condition by performing high-resolution long-term simulations. They found that the strong clumping occurs even for lower dust-to-gas ratios during the time evolution over ten to hundreds Keplerian periods (see their Tables 1 and 2). The critical dust-to-gas ratio at the midplane is about $\simeq0.35-2.5$ in their simulations for $10^{-3}\lesssim\taus\lesssim1$ (see their Figure 4). This may indicate that streaming instability and subsequent clumping operate more easily than previously thought, leading to planetesimal formation.

In such a long-term evolution, dust coagulation may be more important. The coagulation timescale is only a few tens of Keplerian periods for the dust-to-gas surface density ratio of 0.01 \citep[e.g.,][]{Brauer2008}, which can be shorter than the time required for the strong clumping to occur. Besides, the coagulation may be effective even in moderately high density regions around much denser clumps. However, a possible combined process of coagulation and streaming instability is not well studied. We note that there are studies that investigated how disk evolution with coagulation produces a region where streaming instability operates \citep[e.g.,][]{Drazkowska2016,Carrera2017}. What they considered is that dust coagulation and streaming instability occur one after the other: the dust sizes are determined by dust coagulation through the global disk evolution and are independent from the onset of streaming instability and the resulting local turbulence/clumping. In contrast to these studies, we investigate the possible impact of dust growth during the local process of streaming instability, which could change the planetesimal formation efficiency assumed in the previous models. In this paper, 
we demonstrate that dust coagulation considerably proceeds in moderately dense regions or clumps before the strong clumping.

 We consider dust growth toward the size of $\taus=1$. The collision velocity provides the insight into the growth efficiency. When dust grains are so large that the Brownian motion is insignificant, the collision velocity is determined by the drift motion and the turbulent motion. The drift-induced collision velocity of similar-sized dust is on the order of $\taus\eta\vk$, where $\vk$ is the Keplerian velocity, $\eta\sim(\cs/\vk)^2$ is a measure of the radial pressure gradient, and $\cs$ is the gas sound speed \citep[e.g.,][]{Weidenschilling1977}. For $\taus\sim 1$, this velocity is $\sim54\;\mathrm{m/s}$ in the minimum mass solar nebula \citep{Hayashi1981} and lower in a disk that is optically thick to the stellar irradiation \citep[e.g.,][]{Chiang1997}. The turbulence-induced collision velocity depends on $\taus$ and the turbulence strength $\alpha$ \citep[][]{Shakura1973}. When dust grains are large enough, the velocity is on the order of $\sqrt{\alpha\taus}\cs$ for similar-sized grains \citep{Ormel2007}. The collision velocity is thus $\sim10-30\;\mathrm{m/s}$ for $\alpha=10^{-3}$, $\taus=0.1-1$, and $\cs=1\;\mathrm{km/s}$. It has been suggested that such high velocity collisions can lead to fragmentation that limits the dust growth \citep[e.g.,][]{Weidenschilling1993,Blum2008}. Even when the fragmentation is less efficient, the radial drift can limit the dust growth: the drift time is so short that dust grains reach the central star before they grow in size significantly. Before streaming instability operates, we naively expect that dust grains grow to the fragmentation- or drift-limited sizes. We thus address whether or not the clumping due to streaming instability can assist further dust growth.

An advantage of dust coagulation in the clumping regions is lower collision velocity \citep[][]{Johansen2009b,Bai2010b,Schreiber2018}. \citet{Schreiber2018} showed that the local velocity dispersion of the same sized dust grains can be on the order of $10^{-3}\cs$ or much lower in the clumping regions for $\taus=0.1$ (e.g., see Figure 11 therein). \citet{Bai2010b} also showed low collision velocities in the simulations where dust clumping is efficient (their R10Z3 run, see Figures 6 and 10 therein). It is also indicated that large-scale turbulence is ineffective on clump scales once dust grains are locally concentrated \citep[][]{Klahr2020}. This means that, even if the dust growth is limited by fragmentation before the clumping, dense regions shielded from the large-scale turbulence enable further dust growth owing to the lower-speed collisions. Besides, the clumping leads to the reduced drift velocity \citep[see Figures 7 and 8 in][]{Bai2010b}\footnote{ \citet{Bai2010b} also found that the multiple-species effect reduces the drift velocity of each dust species (see solid and dashed (dash-dotted) lines in their Figure 8 for 2D (3D) simulations). We focus on the effect of high dust densities due to clumping (see the panels of R10Z1 and R10Z3 runs as well as Figure 5 therein). }, which is due to the aerodynamical backreaction from dust to gas \citep[][]{Nakagawa1986}. Thus, such dense regions are also preferable for dust grains to overcome the drift barrier. 

This paper is organized as follows. We describe our model in Sections \ref{sec:model} and \ref{sec:collvel}. In Section \ref{sec:model}, we describe the duration time before the strong clumping via streaming instability (a pre-clumping period) and the coagulation timescale. In Section \ref{sec:collvel}, we describe models for dust collision velocities in clumping regions, which is based on the previous simulations of streaming instability. We then compare the coagulation timescale and the pre-clumping period in Section \ref{sec:results}. We show that coagulation proceeds faster than the clumping. In Section \ref{sec:disc_conclusion}, we give conclusions and brief discussion.

\section{Timescale Model}\label{sec:model}

\begin{figure}[tp]
	\begin{center}
	\hspace{100pt}\raisebox{20pt}{
	\includegraphics[width=0.9\columnwidth]{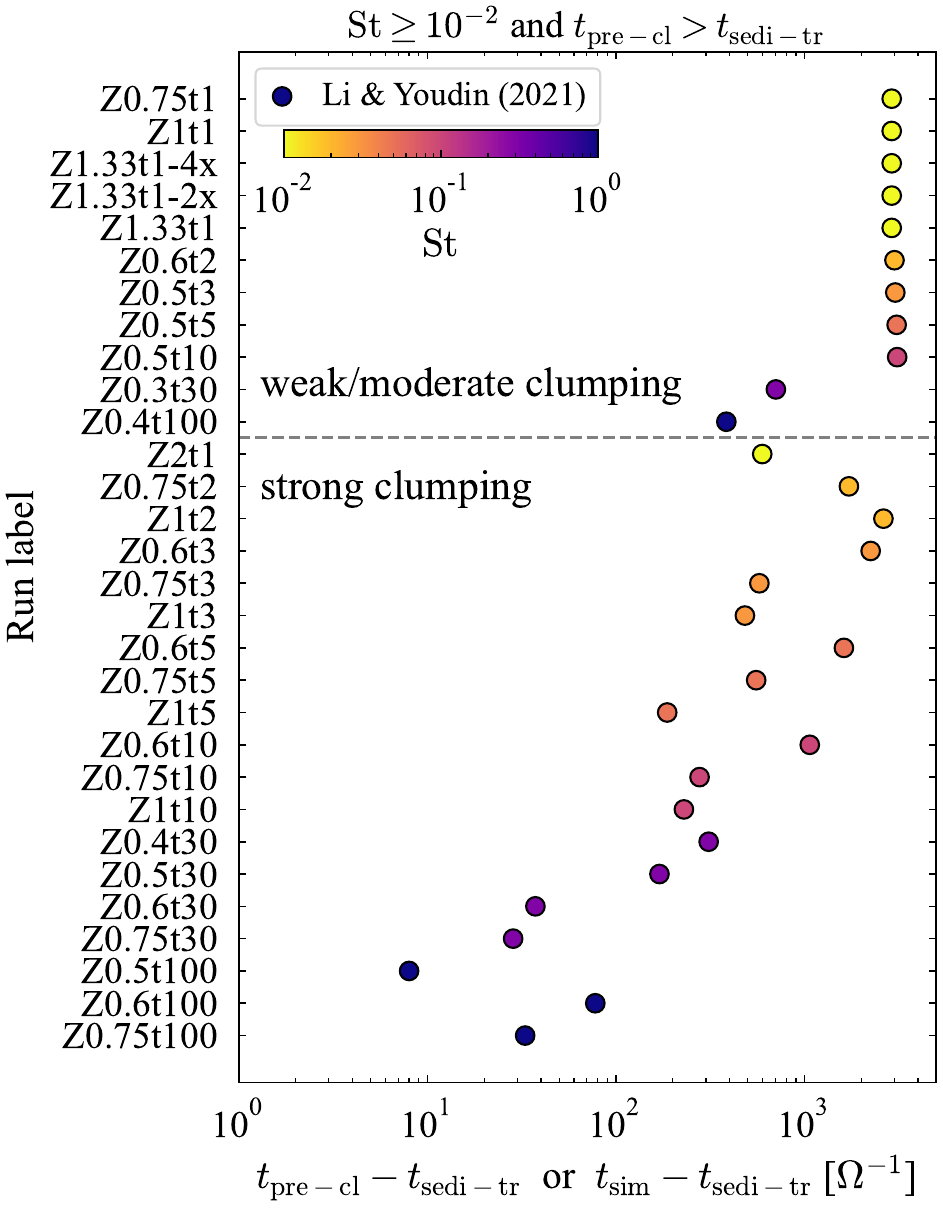} 
	}
	\end{center}
	\vspace{-30pt}
\caption{Duration time of the pre-clumping phase reported in \cite{Li2021}. They define the pre-clumping phase using (1) the time at which the first sedimentation phase ends ($t_{\mathrm{sedi-tr}}$) and (2) the time at which the dust density exceeds $2\rho_{\mathrm{R}}/3$ ($t_{\mathrm{pre-cl}}$). These values are collected from their Table 2. The difference $t_{\mathrm{pre-cl}}-t_{\mathrm{sedi-tr}}$ that we plot is the pre-clumping period. We refer to the simulations that adopt $\taus\geq10^{-2}$ and show $t_{\mathrm{pre-cl}} > t_{\mathrm{sedi-tr}}$. Color of filled circles represents $\taus$ adopted in their simulations \citep[see Table 1 in][]{Li2021}. In this figure, the pre-clumping periods are given by $t_{\mathrm{sim}}-t_{\mathrm{sedi-tr}}$ for weak/moderate clumping cases (see text for details, and see also Table 1 in \citet{Li2021}). \citet{Li2021} also conducted simulations with $\taus=10^{-3}$ (their Z4t0.1, Z3t0.1, and Z2t0.1 runs) and show the pre-clumping periods $\gtrsim475\Omega^{-1}$ (see their Tables 1 and 2), which is longer than $\tauSI\Omega^{-1}$ in our fiducial model. }
\label{fig:LiYoudin2021}
\end{figure}

\subsection{Pre-clumping periods}
We consider a situation where streaming instability has developed into the nonlinear phase and the dust clumping gradually and/or intermittently increases the local dust density as observed in numerical simulations \citep[e.g.,][]{Johansen2007,Johansen2007nature,Bai2010b,Yang2014,Yang2018,Schaffer2018,Xu2022}. As mentioned in the previous section, such evolution is expected when the dust-to-gas mass ratio at the midplane is $\gtrsim 0.35-2.5$ \citep[][]{Li2021}. \citet{Li2021} found the critical dust-to-gas mass ratio of $\simeq0.35-1$ for $\taus\gtrsim0.02$ (see their Figure 4 and the sudden change of the critical value at $0.01<\taus<0.02$). For smaller $\taus$, larger dust-to-gas ratio is required for the strong clumping, and it takes longer time than for $\taus=0.1-1$ according to \citet{Li2021} (see their Tables 1 and 2). Such situations will be more preferable for coagulation to operate before the strong clumping. We thus focus on the dust evolution for $\taus\gtrsim0.01$.

To investigate the possible effect of dust coagulation in clumps before the strong clumping, we estimate the coagulation timescale and compare it to the duration time of the pre-clumping phase (pre-clumping periods). We represent the pre-clumping  period as $\tauSI \Omega^{-1}$, where $\tauSI$ is a numerical factor, and $\Omega$ is the Keplerian angular velocity. In this period, we assume that turbulence driven by streaming instability is nearly saturated and that azimuthally elongated filaments have developed. 

 \citet{Li2021} defined the pre-clumping phase as a phase between the first sedimentation \footnote{ Their simulations show an increase of the dust scale height after the first settling. They include this phase in the first phase that they call the transient sedimentation phase.} ($t<t_{\mathrm{sedi-tr}}$, where $t$ here denotes their simulation time) and the strong clumping phase ($t>t_{\mathrm{pre-cl}}$)\footnote{ \citet{Li2021} define the starting time of the strong clumping phase as the time when the maximum dust density first exceeds $2\rho_{\mathrm{R}}/3$.}. Table 2 in \citet{Li2021} shows $t_{\mathrm{sedi-tr}}$ and $t_{\mathrm{pre-cl}}$ for each run. In Figure \ref{fig:LiYoudin2021}, we plot the pre-clumping period, $t_{\mathrm{pre-cl}}-t_{\mathrm{sedi-tr}}$, for $\taus\geq 0.01$ using the reported values in their table. We refer to the runs that show strong clumping and $t_{\mathrm{pre-cl}}>t_{\mathrm{sedi-tr}}$ (below the gray line). We also refer to their simulations that only show weak/moderate clumping (i.e., $\rho_{\dst,\mathrm{max}}<\rho_{\mathrm{R}}$; above the gray dashed line). For such cases, we plot $t_{\mathrm{sim}}-t_{\mathrm{sedi-tr}}$, where $t_{\mathrm{sim}}$ is the simulation time reported in Table 1 of \citet{Li2021} (e.g., Z0.3t30 run). Although \citet{Li2021} call them non-clumping cases (or weak clumping), the local dust-to-gas ratio increases to $\sim10$ in some of them (e.g., their Z0.3t30 and Z0.4t100 runs; see their interactive figure of the online version of \citet{Li2021}). We thus call them as the moderate clumping in the present paper. Such moderate clumping is enough for dust growth to proceed efficiently, which is shown in Section \ref{sec:results}. 

 According to \citet{Li2021} and Figure \ref{fig:LiYoudin2021}, the pre-clumping period ranges between $\sim10\Omega^{-1} - 1000\Omega^{-1}$. For $\taus<0.3$, the pre-clumping period is greater than $100\Omega^{-1}$.  Although $\tauSI$ should depend on the background $\rhod/\rhog$ and $\taus$, we just assume it to be a constant for simplicity. In this paper, we adopt $\tauSI\Omega^{-1}=100\Omega^{-1}$ as the fiducial pre-clumping period.

\subsection{Dust growth time}

If we assume the perfect sticking, the coagulation timescale is given by
\begin{align}
\tcoag = a\left(\frac{da}{dt}\right)^{-1} &=3m\left(\frac{dm}{dt}\right)^{-1},\\
&=3\frac{m}{\rhod4\pi a^2 \Delta v},
\end{align}
where $a$ and $m$ denote the size and the mass of a spherical dust grain, $\rhod$ is the dust density, $\Delta v$ is the dust-dust collision velocity. Here, we assume equal-mass collisions for simplicity (see also Section \ref{sec:disc_conclusion}). We focus on the mass-dominating dust sizes. If the number density distribution of dust $dn(a)/da$ is proportional to $a^{-3.5}$ or shallower, the mass-dominating dust size roughly corresponds to the largest dust size. Adopting the size distribution of $d\ln n/d\ln a=-3.5$, \citet{Yang2021} conducted numerical simulations of polydisperse streaming instability and found dust segregation that concentrates large dust grains in dense regions. Since we focus on the dust growth in dense regions resulting from streaming instability, the assumption of the equal-mass collisions of the mass-dominating dust grains may be valid. 

In the Epstein regime, the Stokes number, $\taus\equiv\tstop\Omega$, is given as follows:
\begin{equation}
\taus = \sqrt{\frac{\pi}{8}}\frac{\rhoint a}{\rhog \cs}\Omega,
\end{equation}
where $\rhog$ is the gas density. We then rewrite the coagulation timescale in terms of $\taus$:
\begin{equation}
\taucoag\equiv\tcoag\Omega = \sqrt{\frac{8}{\pi}}\frac{\rhog}{\rhod}\frac{\taus \cs}{\Delta v}.\label{eq:tcoag_Ep}
\end{equation}
The coagulation timescale depends on the dust-to-gas ratio $\rhod/\rhog$. An increase in $\rhod/\rhog$ reduces $\taucoag$, i.e. accelerates dust coagulation. If we assume no clumping and that large scale turbulence determines $\Delta v$, one obtains $\taucoag\sim\sigmag/\sigmad$,\footnote{ We also assumed the following: (1) the vertical density distributions of gas and dust are the Gaussian profile, (2) the dust scale height $\hd$ is determined by the large scale turbulence, $\hd\sim H\sqrt{\alpha/\taus}$, where $H\equiv\cs/\Omega$ is the gas scale height \citep[e.g.,][]{Dubrulle1995}, and (3) $\rhod/\rhog$ in Equation (\ref{eq:tcoag_Ep}) is the midplane dust-to-gas ratio.} where $\sigmag$ and $\sigmad$ are the surface densities of gas and dust \citep[e.g.,][]{Brauer2008}.

To take the effect of fragmentation into account, we also utilize the sticking efficiency $p_{\mathrm{eff}}$ introduced in \citet{OH2012} and \citet{Okuzumi2016}:
\begin{equation}
p_{\mathrm{eff}}\equiv \mathrm{min}\left(1, -\frac{\ln\left(\Delta v/\vfrag\right)}{\ln 5}\right),\label{eq:peff}
\end{equation}
where $\vfrag$ is the critical fragmentation velocity. We then scale the coagulation timescale with the sticking efficiency as follows:
\begin{equation}
\taucoageff = \taucoag/p_{\mathrm{eff}}.\label{eq:tcoag_eff}
\end{equation}
The critical fragmentation velocity has been investigated by both laboratory experiments and numerical simulations \citep[e.g.,][]{Blum2000,Blum2008,Wada2009,Wada2013,Hasegawa2021}. However, the value of $\vfrag$ still seems uncertain. We thus adopt $\vfrag=10\;\mathrm{m/s}$ as a fiducial case (see also Section \ref{sec:disc_conclusion}). This value has been used in the model treating coagulation and planetesimal formation due to streaming instability \citep[e.g.,][]{Drazkowska2016}.

\section{Collision velocity model}\label{sec:collvel}
Once we model $\Delta v$ as a function of dust properties, we can estimate how efficiently coagulation proceeds using Equation (\ref{eq:tcoag_eff}). As a simple model, we consider the following form of the collision velocity:
\begin{equation}
\Delta v \propto \cs \taus^{A}(\rhod/\rhog)^B,
\end{equation}
where $A$ and $B$ are constant. We assume the velocity dispersion of dust grains to be $\Delta v$ in this work. The velocity dispersion has been measured in the previous numerical simulations \citep[e.g.,][]{Johansen2009b,Bai2010b,Schaffer2018,Schreiber2018,Yang2021}. We follow \citet{Schreiber2018} and \citet{Bai2010b} in this work since (1) \citet{Schreiber2018} showed the dependence of the velocity dispersion on the dust-to-gas ratio in small-domain simulations (e.g., see Figure 5 therein), and (2) \citet{Bai2010b} showed the velocity dispersion for larger $\taus$ with the $\rhod/\rhog$-dependence (e.g., see Figure 13 therein).
 
 \subsection{Model 1: \citet{Schreiber2018}}\label{subsec:model1}
\citet{Schreiber2018} conducted 2D simulations with $\taus=0.01$ and 0.1 and with the background dust-to-gas ratio ranging from 0.1 to $10^3$. Some of their simulations adopted very small domain size of $10^{-3}H$, where $H$ is the gas scale height. Velocity dispersion measured in such a small domain is suitable to discuss dust collisions at small spatial scales. They derived velocity dispersion with respect to the domain-averaged velocity and one with respect to the cell-averaged velocity. The former could include relative velocities between individual dust clumps. We refer to the latter, which seems to be more important for dust collisions. We note that their simulations treat a single dust species, and thus the measured velocity dispersion is for the same dust sizes. This is sufficient for the purpose of this work since the previous coagulation simulations showed that the dust growth is dominated by similar-sized dust collisions \citep[e.g.,][]{Okuzumi2012}.

They found that the velocity dispersion is almost constant for $1\lesssim\rhod/\rhog\lesssim 10$, and $\sim10^{-3}\cs$ for $\taus=0.1$. On the other hand, the velocity dispersion is proportional to $(\rhod/\rhog)^{-1}$ for $\rhod/\rhog\gtrsim10$ (Figures 5, 11, 16 and 21 therein). As for the $\taus$-dependence, their simulations show a relatively weak dependence with $0<A<1$ (e.g., see their Figures 11 and 21 for the runs with the radial and vertical dimensions). Based on these results, we model the velocity dispersion ($\Delta v=\Delta v_1$) as follows:
\begin{equation}
\Delta v_1 =
\begin{cases}
\displaystyle 10^{-3}\cs \left(\frac{\taus}{10^{-1}}\right)^{0.5} &\displaystyle \left(\frac{\rhod}{\rhog}<10\right)\\
\displaystyle10^{-3}\cs \left(\frac{\taus}{10^{-1}}\right)^{0.5}\left(\frac{\rhod/\rhog}{10}\right)^{-1}. &\displaystyle \left(\frac{\rhod}{\rhog}\geq10\right)
\end{cases}
\label{eq:dv_model1}
\end{equation}
 
This collision velocity $\Delta v_1$ is lower than the maximum drift speed $\eta \vk$:
\begin{equation}
\eta \equiv -\frac{1}{2}\left(\frac{\cs}{\vk}\right)^2\frac{d\ln P}{d \ln r},
\end{equation}
where $P$ is the gas pressure and $r$ is the radial distance from the central star \citep[e.g.,][]{Adachi1976,Weidenschilling1977}. The maximum drift velocity is thus $\eta\vk\sim (\cs/\vk)\times\cs$. The disk aspect ratio $\cs/\vk$ is $\sim10^{-2}-10^{-1}$ \citep[e.g.,][]{DAlessio1999}. Thus, $\Delta v_1$ is about ten times lower than the maximum drift speed. This collision velocity becomes even lower in the clumping regions of $\rhod/\rhog>10$. This low collision velocity is favorable for dust growth without significant fragmentation.

\begin{figure}[tp]
	\begin{center}
	\hspace{100pt}\raisebox{20pt}{
	\includegraphics[width=0.9\columnwidth]{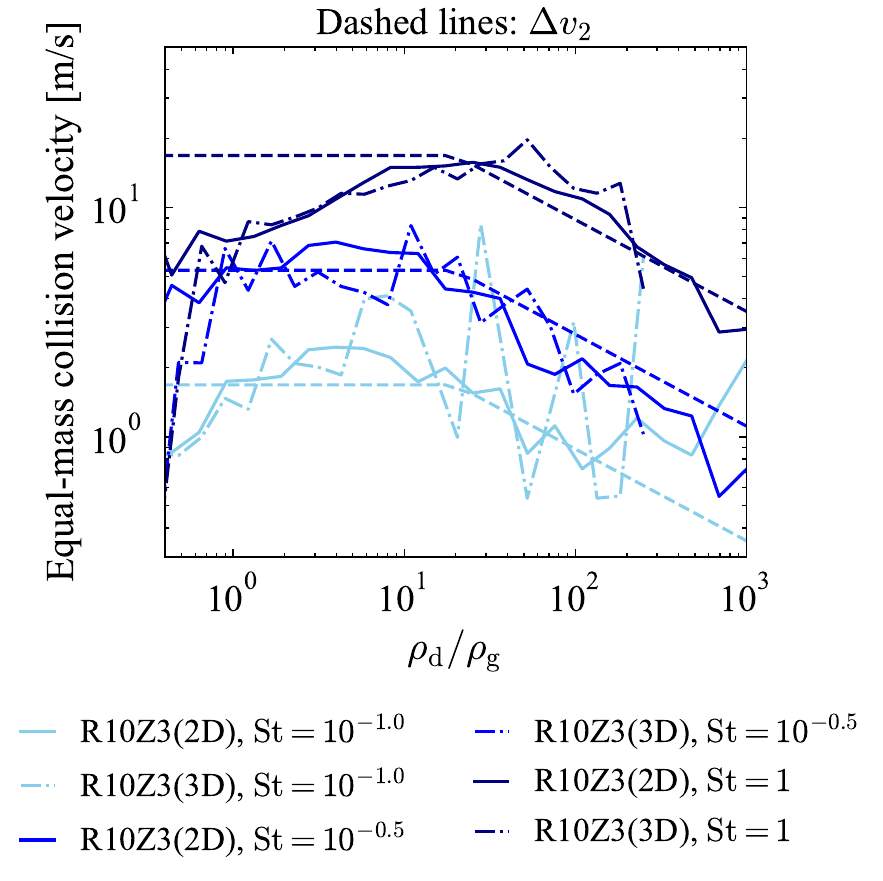} 
	}
	\end{center}
	\vspace{-30pt}
\caption{Comparison of our collision velocity model $\Delta v_2$ and the simulation results of \citet{Bai2010b} ($\cs=0.99\;\mathrm{km/s}$). Solid and dash-dotted lines show the collision velocities measured in the R10Z3 run of \citet{Bai2010b} with 2D and 3D domains, respectively (see their Figure 13). Dashed lines show our model (Equation (\ref{eq:dv_model2})). We show the collision velocities of each $\taus$ with different color.  } 
\label{fig:BS10_model}
\end{figure}

\begin{figure*}[tp]
	\begin{center}
	\hspace{0pt}\raisebox{20pt}{
	\includegraphics[width=1.6\columnwidth]{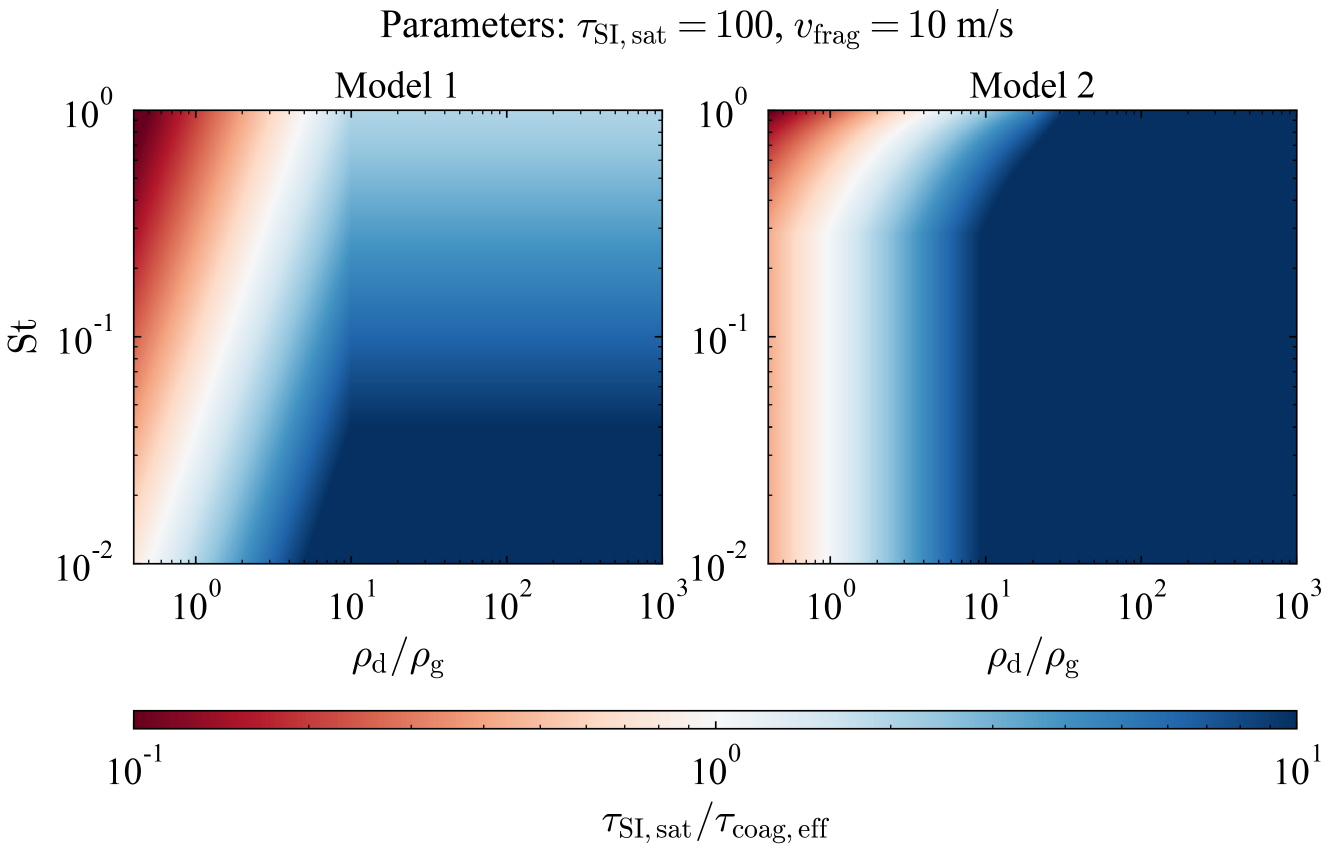} 
	}
	\end{center}
	\vspace{-30pt}
\caption{Ratio of the pre-clumping period $\tauSI$ and the coagulation timescale $\taucoageff$. We assume the gas temperature of $50\;\mathrm{K}$ ($\cs\simeq0.4\;\mathrm{km/s}$) as an example. We note that we take into account the effect of the possible imperfect sticking (Equation (\ref{eq:tcoag_eff})). The blue (red) region indicates that the growth time is shorter (longer) than the pre-clumping period before the strong clumping. We naively expect that the dust evolves along the white region ($\tauSI\sim\taucoageff$) on this $\rhod/\rhog-\taus$ plane (see also Figure \ref{fig:schematic}).} 
\label{fig:tratio}
\end{figure*}

 \subsection{Model 2: \citet{Bai2010b}}\label{subsec:model2}
The simulations in \citet{Schreiber2018} are limited to the cases of $\taus=0.01$ and 0.1. As shown in the subsequent section, the clumping due to streaming instability can promote further dust growth: dust grains accumulated locally in a clump collide with each other, and their Stokes number will increase further. Thus, in addition to Model 1, we also refer to \citet{Bai2010b} who performed multiple-dust-size simulations with larger dust grains. The simulations of \cite{Bai2010b} also treats the vertical stratification in 2D and 3D domains while the simulations of \citet{Schreiber2018} considered unstratified disks with smaller domains.

They assumed the uniform dust mass distribution over their assumed size ranges (see Section 2.3 therein). Among their simulations with different size ranges and total metallicities, we refer to the R10Z3 run, where the size range is $0.1\leq\taus\leq 1$ and the metallicity is 0.03. The reasons for referring to this run are as follows. First, the previous studies on dust coagulation in a disk show that the dust mass distribution tends to be top-heavy \citep[e.g.,][]{Brauer2008,Birnstiel2012,Okuzumi2012}. The uniform mass distribution in the numerical simulation of streaming instability would be valid if the size range is narrow and covers the mass-dominating sizes, as noted in \citet{Bai2010b}. The R10Z3 run is one of the runs with the narrowest size range in the logarithmic space. Second, the focus of our study is the dust coagulation in the pre-clumping phase, and thus we need the collision velocity in dust dense regions. The R10Z3 run shows efficient dust concentration in both 2D and 3D simulations while the other runs labeled R21Z3 and R30Z3 show very small volume fraction of the dust dense region in 3D (see Figure 6 therein). The high volume fraction of the dust dense region in the R10Z3 run will ensure better statistics of the collision velocity. 

The dust collision velocity in the R10Z3 run is shown in their Figure 13. We focus on the equal mass collision for clear comparison with Model 1. According to their results, the equal mass collision velocity is larger for larger dust (Figure \ref{fig:BS10_model}). As in \citet{Schreiber2018}, \citet{Bai2010b} also show the decrease of the collision velocity in the dust-dense region. Based on their data, we construct the following model as Model 2 ($\Delta v=\Delta v_2$):
\begin{equation}
\Delta v_2 =
\begin{cases}
\displaystyle 1.7\times10^{-2}\cs \left(\frac{\taus}{1}\right) &\displaystyle \left(\frac{\rhod}{\rhog}<20\right)\\
\displaystyle 1.7\times10^{-2}\cs \left(\frac{\taus}{1}\right)\left(\frac{\rhod/\rhog}{20}\right)^{-0.4}. &\displaystyle \left(\frac{\rhod}{\rhog}\geq20\right)
\end{cases}
\label{eq:dv_model2}
\end{equation}
We derive the prefactor of $1.7\times10^{-2}$ and the $\rhod/\rhog$-dependence by fitting $\Delta v_2/\taus$ to the collision velocities of $\taus=10^{-0.5}$ and $1$ in the high-density regions of $\rhod/\rhog>20$. We note that their simulations show higher abundance of particles of $\taus=10^{-0.5}$ and $1$ than particles of $\taus=10^{-1}$ \citep[see Figure 6 of][]{Bai2010b}, which may indicate better statistics for larger particles. We thus use the larger particle data for the fitting. In Figure \ref{fig:BS10_model}, we compare $\Delta v_2$ and the simulation data of \citet{Bai2010b} for $\taus=10^{-1},\; 10^{-0.5}$, and $1$. Our model underestimates the collision velocity for $\taus=10^{-1}$ and $\rhod/\rhog>10^2$. Thus, the collisional growth of dust particles of $\taus=10^{-1}$ is more efficient than our model predicts unless the critical fragmentation velocity is low. Our model overestimates the collision velocity of $\taus=1$ for $\rhod/\rhog\lesssim10$, and thus the coagulation efficiency is overestimated by a factor of a few. The collision velocity of $\taus=10^{-0.5}$ is better represented by our model.

The collision velocity $\Delta v_2$ is greater than $\Delta v_1$, which may be due to the vertical motion in the stratified disk. The collision velocity can be comparable to the maximum drift velocity for $\rhod/\rhog<20$ or only slightly lower. Nevertheless, the dust enhancement of $\rhod/\rhog>20$ reduces the collision velocity and makes fragmentation less efficient. This is thus favorable for dust growth \citep[see also][]{Johansen2009b}.

\begin{figure}[tp]
	\begin{center}
	\hspace{100pt}\raisebox{20pt}{
	\includegraphics[width=0.8\columnwidth]{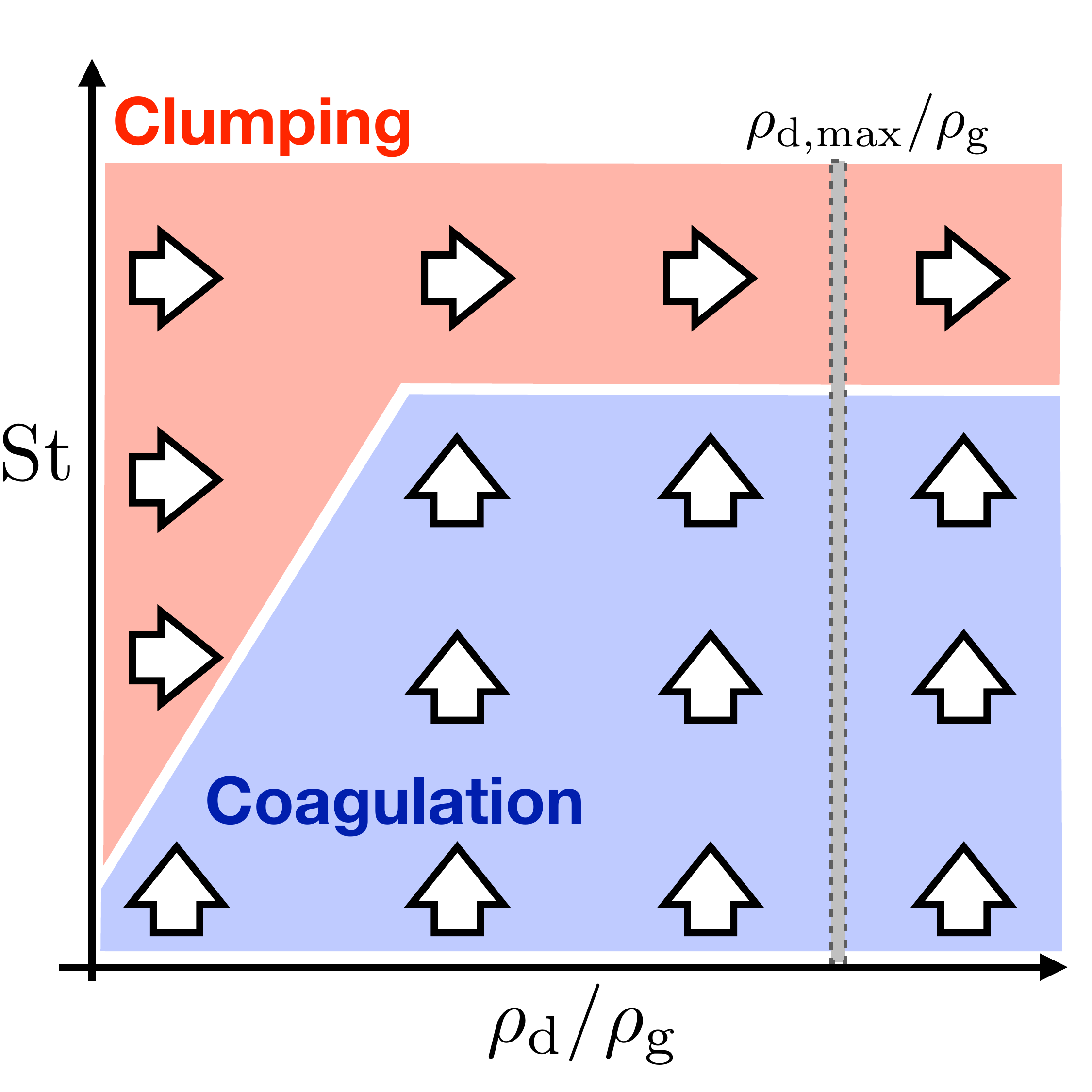} 
	}
	\end{center}
	\vspace{-30pt}
\caption{Schematic figure to show dust evolution on the $\rhod/\rhog-\taus$ plane. We assume Model 1 for the collision velocity as an example. The dust moves upward if the coagulation timescale is shorter than $\tSI$ (the blue region). For $\tauSI < \taucoageff$, the clumping increases the dust density and thus the dust moves rightward on the $\rhod/\rhog-\taus$ plane (the red region). Therefore, we naively expect that the dust first moves to the boundary of the blue and red regions and then moves along it, which is independent from initial values of $\taus$ and $\rhod/\rhog$. The gray region indicates the maximum density that can be expected via the strong clumping due to streaming instability.} 
\label{fig:schematic}
\end{figure}

\section{Results}\label{sec:results}
\subsection{Dust evolution on $\rhod/\rhog-\taus$ plane}
Figure \ref{fig:tratio} shows $\tauSI/\taucoageff$ as a function of $\rhod/\rhog$ and $\taus$. The collision velocity used to plot the left panel is the Model 1 while that for the right panel is the Model 2. We assume the gas temperature of $50\;\mathrm{K}$ and thus $\cs\simeq0.4\;\mathrm{km/s}$ as an example. We note that the gas temperature affects the coagulation timescale only through $p_{\mathrm{eff}}$ (see also Equation (\ref{eq:tcoag_Ep})). 
We find that, for the high dust-to-gas ratio, coagulation proceeds before the strong clumping ($\tauSI>\taucoageff$). The strong clumping dominates over coagulation when $\rhod/\rhog$ is relatively small or $\taus$ is sufficiently large. We note that, regardless of low collision velocities, the dust growth timescale in clumps is comparable to or shorter than the growth timescale in the non-clumping case where $\tcoag\sim100\Omega^{-1}$ for $\sigmad/\sigmag=0.01$ \citep[][]{Brauer2008}.

One can see how the dust properties evolve with time on the $\rhod/\rhog-\taus$ plane (see also Figure \ref{fig:schematic}). In the blue region ($\tauSI>\taucoageff$), the coagulation is more efficient than the dust clumping. This means that $\taus$ increases faster than $\rhod/\rhog$, and thus the dust evolution is in the vertical direction and upward on the $\rhod/\rhog-\taus$ plane. On the other hand, in the red region, the dust clumping is faster than the coagulation, which means the efficient increase of $\rhod/\rhog$. The dust evolution is then in the horizontal direction and rightward. In this way, we naively expect that the dust moves first to the critical line of $\tauSI\sim\taucoageff$ and then roughly along it, which is independent from initial values of $\taus$ and $\rhod/\rhog$.

In the case of the Model 1 (the left panel of Figure \ref{fig:tratio}), the critical line is inclined for $\rhod/\rhog<10$, and thus both coagulation and clumping appear to proceed. Once the dust-to-gas ratio becomes larger than 10, the coagulation dominates over the clumping for $10^{-2}\leq\taus\leq 1$. The coagulation timescale in this regime is given by
\begin{equation}
\taucoageff =p_{\mathrm{eff}}^{-1}\sqrt{\frac{8}{\pi}}10^{1.5}\sqrt{\taus},
\end{equation}
which is independent from $\rhod/\rhog$. Thus, the coagulation timescale is comparable to $\tauSI$ for the following stopping time:
\begin{equation}
\taus = p_{\mathrm{eff}}^2\frac{\pi}{8}10^{-3}\tauSI^2\simeq 4p_{\mathrm{eff}}^2\left(\frac{\tauSI}{100}\right)^2.\label{eq:taus_crit}
\end{equation}
For the perfect sticking case ($p_{\mathrm{eff}}=1$), the clumping dominates over the coagulation for $\taus>0.36$ and $\taus>4$ when $\tauSI$ is 30 and $100$, respectively. In this region, the dust density increases efficiently (see also Figure \ref{fig:schematic}).

In the case of the Model 2, the coagulation timescale is comparable to $\tauSI=100$ for $\rhod/\rhog\sim1$ when dust grains are relatively small ($\taus\lesssim0.3$). Thus, coagulation dominates the dust evolution, and the dust moves along the vertical white region. As the dust becomes larger ($\taus\gtrsim0.3$), the collision velocity becomes high, and collisions lead to the imperfect sticking. As a result, the clumping dominates for $\rhod/\rhog\sim1$ and $\taus\gtrsim0.3$, leading to an increase in the dust density. For high dust-to-gas ratios ($\rhod/\rhog>10$), the coagulation again becomes faster than the clumping. Therefore, we expect coagulation to be important for Model 2 as well. \citet{Bai2010b} also discussed the combined effect of coagulation and the clumping, indicating the positive feedback between them (see Section 6 therein). Our results are consistent with their discussion.

\begin{figure}[tp]
	\begin{center}
	\hspace{100pt}\raisebox{20pt}{
	\includegraphics[width=0.9\columnwidth]{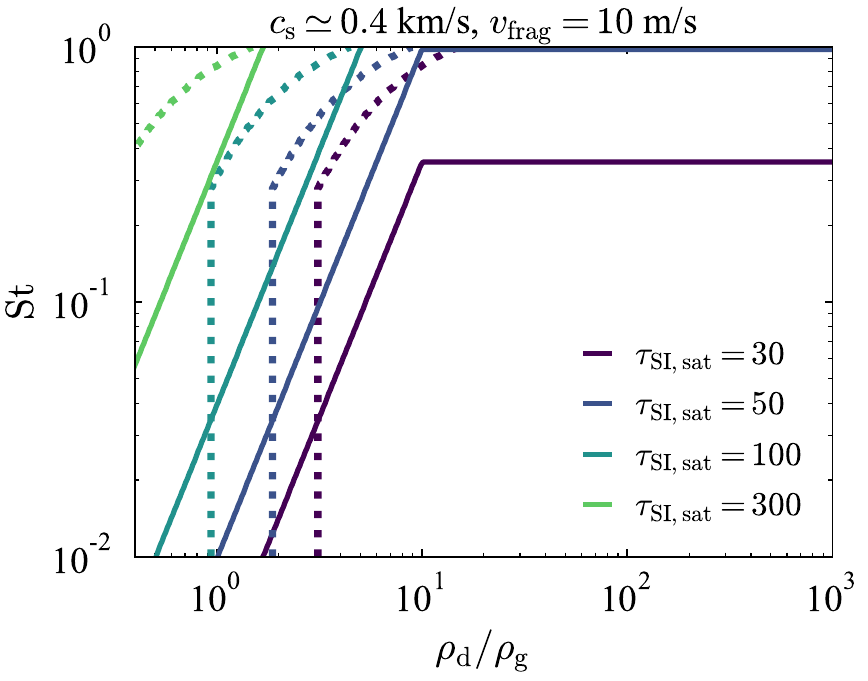} 
	}
	\end{center}
	\vspace{-30pt}
\caption{Dependence of the critical line ($\tauSI=\taucoageff$) on $\tauSI$. The solid and dotted lines represent cases where we adopt Model 1 and Model 2, respectively. As in Figure \ref{fig:tratio}, we assume the gas temperature of 50 K to plot this figure. The critical $\taus$ for which the solid line becomes horizontal (see the case of $\tauSI=30,\; 50$) is given by Equation (\ref{eq:taus_crit}).} 
\label{fig:tsi_depend}
\end{figure}

\begin{figure*}[tp]
	\begin{center}
	\hspace{0pt}\raisebox{20pt}{
	\includegraphics[width=1.5\columnwidth]{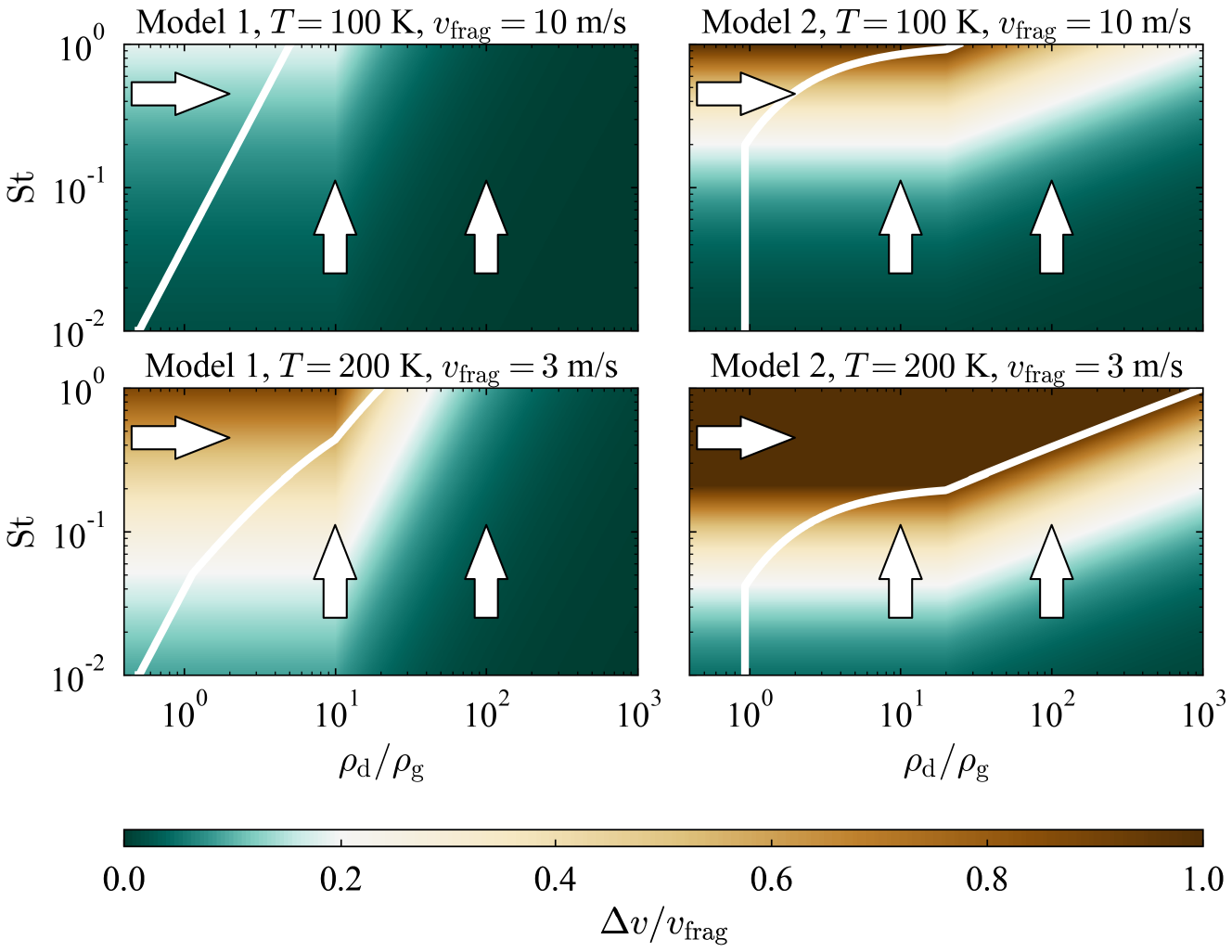} 
	}
	\end{center}
	\vspace{-30pt}
\caption{Collision velocities as a function of $\taus$ and $\rhod/\rhog$ for $T=100\;\mathrm{K}$ and $200\;\mathrm{K}$, for which $\cs$ is $\simeq0.6\;\mathrm{km/s}$ and $\simeq0.8\;\mathrm{km/s}$, respectively. The adopted critical fragmentation velocity in each case is $\vfrag=10\;\mathrm{m/s}$ and $\vfrag=3\;\mathrm{m/s}$, which is based on the fact that silicate grains are often assumed to be more fragile than water ice \citep[but see also][]{Kimura2015,Steinpilz2019}. The white line marks the critical line ($\tauSI=\taucoageff$). The arrows show the direction of the dust evolution on $\rhod/\rhog-\taus$ plane. Dust grains suffer fragmentation in a wider parameter space than in Figure \ref{fig:tratio} because of higher temperature and smaller $\vfrag$. Nevertheless, dust growth time is shorter than the pre-clumpting period space once the local $\rhod/\rhog$ increases beyond 15 in the Model 1 for $T=200\;\mathrm{K}$ (20 in the Model 2 for $T=100\;\mathrm{K}$).
In the Model 2 with $T=200\;\mathrm{K}$ (the bottom right panel), the critical line is close to the fragmentation-limited line ($\Delta v=\vfrag$) for $\taus\gtrsim0.2$. } 
\label{fig:tratio_silic}
\end{figure*}

\begin{figure}[tp]
	\begin{center}
	\hspace{100pt}\raisebox{20pt}{
	\includegraphics[width=0.9\columnwidth]{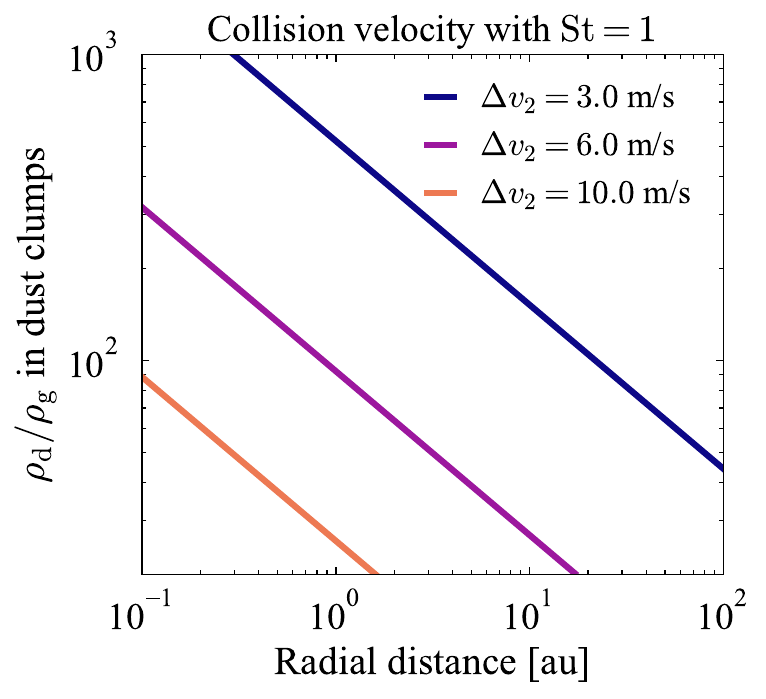} 
	}
	\end{center}
	\vspace{-30pt}
\caption{ Isolines of the collision velocity of the Model 2 $\Delta v_2$ with $\taus=1$ on $r-\rhod$ plane. We adopt $120\;\mathrm{K} (r/1\;\mathrm{au})^{-3/7}$ \citep[][]{Chiang2010} and plot lines of $3\;\mathrm{m/s},\;6\;\mathrm{m/s}$ and $10\;\mathrm{m/s}$. Note that the collision velocity decreases as $\rhod/\rhog$ increases for $\rhod/\rhog\geq20$ (Equation (\ref{eq:dv_model2})). }
\label{fig:dv2_as_r_and_d2g}
\end{figure}

 In Figure \ref{fig:tsi_depend}, we show the $\tauSI$-dependence of the critical line ($\tauSI=\taucoageff$) for both Model 1 (solid lines) and Model 2 (dotted lines). As shown by Equation (\ref{eq:taus_crit}), the critical line for the Model 1 becomes horizontal at $\taus\simeq0.36$ and $\taus=1$ for $\tauSI=30$ and $50$, respectively. Thus, the clumping takes over the coagulation earlier in the smaller-$\tauSI$ cases. In the Model 2, larger $\tauSI$ leads to wider ranges of $\taus$ where coagulation dominates the clumping (see the regions where the dotted lines are inclined). Thus, even in the presence of the imperfect sticking, coagulation can be the mechanism to govern the dust evolution.

According to \citet{Schreiber2018}, the dust-to-gas ratio increases by a factor of $\sim$10-100 from the background value via the dust clumping in the nonlinear phase of turbulence due to streaming instability. Other studies also show a similar increase \citep[e.g.,][]{Yang2014,Yang2017}. Although \citet{Li2021} show that the maximum dust density $\rho_{\dst,\mathrm{max}}$ can be $\sim10^3\rhog$ in their simulations, the dust-to-gas ratio has a time variation and oscillates in $\sim10^2\rhog-10^3\rhog$ (e.g., see Figures 3 and 12 therein). Thus, it would be reasonable to expect $\rho_{\dst,\mathrm{max}}/\rhog$ to be $\sim10^2$ on average if the clumping starts at $\rhod/\rhog\sim1$. This means that, even if the dust moves to the clumping regime (e.g., the red region in Figure \ref{fig:schematic}), the dust density stops increasing around $\rho_{\dst,\mathrm{max}}$ (the gray region in Figure \ref{fig:schematic}) unless many resulting filaments merge so that the density increases further \citep[see also][for the dependence of the maximum density on the mass reservoir]{Li2018}.

The fate of the dust clumps depends on whether or not $\rho_{\dst,\mathrm{max}}$ is greater than the Roche density $\rho_{\mathrm{R}}\equiv9\Omega^2/4\pi G$, where $G$ is the gravitational constant. On the one hand, the dust clumps have the potential to collapse self-gravitationally for $\rho_{\dst,\mathrm{max}}>\rho_{\mathrm{R}}$. For the self-gravitational collapse, it is also required that the self-gravity of the dust clump dominates over the turbulent diffusion \citep[e.g.,][]{Gerbig2020,Klahr2020,Klahr2021}. \citet{Klahr2020} found that dust clumps larger than $60-120\;\mathrm{km}$ can collapse, which is based on the solar nebula model of \citet{Lenz2020}.  On the other hand, for $\rho_{\dst,\mathrm{max}}<\rho_{\mathrm{R}}$, the dust evolves only through coagulation. Although the coagulation in such a regime takes a longer time than $\tauSI\Omega^{-1}$, the large dust-to-gas ratio leads to lower collision velocity, and dust grains will avoid significant fragmentation \citep[see also][for a similar process]{Tominaga2022b}. Besides, the dust drift becomes slower for $\taus>1$, which also helps dust grains to grow beyond the drift barrier. 

In the above estimate, we assume $T=50\;\mathrm{K}$, and thus the results are applicable in the context of icy planetesimal formation. To investigate the temperature dependence, we also estimate $\tauSI/\taucoageff$ and the collision velocity for $(T,\vfrag) = (100\;\mathrm{K},\;10\;\mathrm{m/s})$ and $(200\;\mathrm{K},\;3\;\mathrm{m/s})$. We adopt smaller $\vfrag$ in the case of $T=200\;\mathrm{K}$ since silicate grains are often assumed to be more fragile than water ice \citep[but see also][]{Kimura2015,Steinpilz2019}. Figure \ref{fig:tratio_silic} shows the critical lines and the collision velocities. In the case of the Model 1 with $T=100\;\mathrm{K}$ (the top left panel), the critical line is the same as in Figure \ref{fig:tratio}, since $p_{\mathrm{eff}}$ is unity in the plotted region and $\taucoag$ is independent of $\cs$ (see Equations (\ref{eq:tcoag_eff}) and (\ref{eq:dv_model1})). In the case of the Model 2 (the top right panel), the region where the imperfect sticking occurs is larger than in Figure \ref{fig:tratio}. Nevertheless, once $\rhod/\rhog$ increases beyond $\simeq 20$, the coagulation dominates over the clumping even for large dust of $\taus\sim1$.

In the cases of $(T,\vfrag)=(200\;\mathrm{K},\;3\;\mathrm{m/s})$, dust grains suffer fragmentation even with the Model 1 since $\Delta v_1$ becomes high (e.g., $\simeq 1\;\mathrm{m/s}$ for $\taus>0.1$ and $\rhod/\rhog<10$; see the bottom left panel). The dust growth is thus delayed compared to the perfect sticking case. Nevertheless, the collision velocity is still lower than 3 m/s by a factor of $\lesssim0.6$ (e.g., $\Delta v_1/\vfrag\simeq0.6$ for $\rhod/\rhog=10$ and $\taus=0.5$; see Equation (\ref{eq:dv_model1})). For the Model 2, the collision velocity is very close to $\vfrag$ for $\taus\gtrsim0.2$ along the critical line (the bottom right panel). Fragmentation significantly slows the dust growth down, and thus the increase in $\rhod/\rhog$ via streaming instability dominates over coagulation in a wider parameter space than in the top panels and in Figure \ref{fig:tratio}. If the strong clumping or the merger of filaments leads to $\rhod/\rhog\sim10^3$, silicate dust grains can grow toward $\taus=1$ for the Model 2 (see Section \ref{subsec:fragStSI}). 

 We also estimate the drift timescale during the dust evolution along the white line in Figure \ref{fig:tratio_silic}. The drift timescale $\tdri$ is given by
\begin{equation}
\tdri \equiv \frac{r}{|\vdri |}=\frac{\left(1+\epsilon\right)^2+\taus^2}{2\taus\eta\Omega},
\end{equation}
where we use the drift velocity $\vdri$ that depends on both $\taus$ and $\epsilon\equiv\rhod/\rhog$ \citep{Nakagawa1986}. The timescale of the dust evolution along the white line is on the order of $\tauSI\Omega^{-1}$ since the coagulation timescale is comparable to the clumping timescale. Streaming instability and the coagulation in clumps will be ineffective if $\tdri$ is smaller than $\tauSI\Omega^{-1}$ from the beginning. We thus consider $\tdri>\tauSI\Omega^{-1}$ in the initial state $(\epsilon,\;\taus)=(\epsilon_0,\;\mathrm{St}_0)$. Assuming the change in $\eta$ to be insignificant compared to $\taus$ and $\rhod/\rhog$, we obtain an increasing factor of the drift timescale with respect to the initial value:
\begin{align}
\frac{\tau_{\mathrm{drift}}(\epsilon,\;\taus)}{\tau_{\mathrm{drift}}(\epsilon_0,\;\mathrm{St}_0)}&=\frac{\mathrm{St}_0}{\taus}\frac{\left(1+\epsilon\right)^2+\taus^2}{\left(1+\epsilon_0\right)^2+\mathrm{St}_0^2},\notag\\
&\sim \frac{\mathrm{St}_0}{\taus}\left(\frac{1+\epsilon}{1+\epsilon_0}\right)^2.
\end{align}
where $\tau_{\mathrm{drift}}\equiv\tdri\Omega$ and we assumed $(1+\epsilon)^2\gg\taus^2$ and $(1+\epsilon_0)^2\gg\mathrm{St}_0^2$ considering the dust evolution along the white lines in Figure \ref{fig:tratio_silic}. In the case of the top left panel of Figure \ref{fig:tratio_silic}, the white line shows $\taus\propto\epsilon^2$, and the drift timescale is almost constant along the white line. The slopes of the white lines on the other panels are shallower for $\Delta v/\vfrag>0.2$, meaning $\tau_{\mathrm{drift}}(\epsilon,\;\taus)>\tau_{\mathrm{drift}}(\epsilon_0,\;\mathrm{St}_0)$ for $\epsilon>\epsilon_0$. Therefore, the drift speed of dust grains does not increase during their growth in clumps and the clumping. The clumping can help dust grains to overcome the drift barrier \citep[see also][]{Bai2010b}.\footnote{ Because of the $r$-dependence of $\Omega$, $\tdri(\epsilon,\;\taus)$ can be shorter than $\tdri(\epsilon_0,\;\mathrm{St}_0)$. However, the coagulation timescale and the clumping timescale also scale with $\Omega^{-1}$ \citep[see][for the coagulation timescale]{Brauer2008}. We thus ignore the factor of $\Omega$. }

\subsection{ On the possible fragmentation limit during streaming instability}\label{subsec:fragStSI}
 For the Model 1, we find that fragmentation affects the dust growth, but it just moderately slows the dust growth down. The collision velocity of the Model 2 can be, however, very close to $\vfrag=3\;\mathrm{m/s}$ even for $\rhod/\rhog\gg 10$. This indicates that, in the case of the Model 2, efficient fragmentation regulates or limits the dust growth unless the clumping increases $\rhod/\rhog$ sufficiently.

 To see where and when the fragmentation prevents the dust growth beyond $\taus=1$, we adopt a simple temperature profile, $T=120\;\mathrm{K}(r/1\;\mathrm{au})^{-3/7}$ \citep[][]{Chiang2010}, and see the $r-$dependence of the collision velocity of the Model 2. Since the velocity also depends on $\rhod/\rhog$, we assume $\taus=1$ and plot isolines of the collision velocity. From Equation (\ref{eq:dv_model2}) with $\rhod/\rhog\geq 20$ and $\taus=1$, one gets a formula of an isoline, $\Delta v_2(\taus=1)=\Delta v$,
\begin{equation}
\frac{\rhod}{\rhog} \simeq 75\left(\frac{\cs}{1\;\mathrm{km/s}}\right)^{2.5}\left(\frac{\Delta v}{10\;\mathrm{m/s}}\right)^{-2.5}. \label{eq:d2g_for_dv}
\end{equation}

 In Figure \ref{fig:dv2_as_r_and_d2g}, we plot three isolines of the collision velocity of the Model 2 $\Delta v_2(\taus=1)$ on the $r-\rhod$ plane. We note that the collision velocity decreases as $\rhod/\rhog$ increases, and also note that the vertical axis of Figure \ref{fig:dv2_as_r_and_d2g} represents the dust-to-gas ratio in dust clumps. For the adopted temperature model, the collision velocity can be lower than 3 m/s at $r\lesssim0.5\;\mathrm{au}$ ($T\gtrsim160\;\mathrm{K}$) if the dust-to-gas ratio in a clump exceeds $\sim 800$. If $\vfrag$ is $\simeq 6\;\mathrm{m/s}$ for silicate dust with 0.1 $\mu\mathrm{m}$-sized monomers \citep{Wada2009}, the required $\rhod/\rhog$ is about five times smaller (see also Equation (\ref{eq:d2g_for_dv})). Growth of silicate dust grains is even easier if their surface energy is ten times higher than previously assumed, which is indicated in \citet{Kimura2015} and \citet{Steinpilz2019}.

 For the temperature profile we adopted, the collision velocity of the Model 2 for $\rhod/\rhog\geq20$ is lower than 10 m/s beyond $\simeq 1.6\;\mathrm{au}$. Thus, regardless of higher collision velocity than the Model 1, the growth of icy dust grains will not be inhibited in the turbulent state due to streaming instability.

\section{Conclusions and Discussion}\label{sec:disc_conclusion}
Based on the previous numerical studies of streaming instability, we model the collision velocity and compare the coagulation timescale and the duration time of the pre-clumping phase (the pre-clumping period). We show that even moderately increased dust density due to streaming instability promotes dust coagulation (Figure \ref{fig:tratio}). It is expected that dust evolves roughly along the line of $\taucoageff\sim\tauSI$ on the $\rhod/\rhog-\taus$ plane (Figure \ref{fig:schematic}). The combination of dust growth and the clumping might allow dust growth toward and beyond $\taus=1$ if the clumping leads to sufficient deceleration of dust drift. Our results highlight the importance of numerical simulations that consider both coagulation and streaming instability. 

 Once dust grains grow beyond $\taus=1$ via the rapid coagulation, streaming instability becomes inefficient to sustain the clumping state because of weak dust-gas coupling. Such large dust grains will then settle toward the midplane, gradually increasing the midplane dust density. If a sufficient amount of large dust grains form and settle, planetesimal formation via gravitational instability will occur \citep[e.g.,][]{Michikoshi2010,Michikoshi2017}. If the increase of the midplane dust density is too slow, the evolution of the large dust grains might be regulated by erosion or mass transfer through collisions with remaining smaller dust grains \citep[][]{Krijt2015,Hasegawa2021}.

One important parameter in this study is the pre-clumping period $\tauSI$. Although we assume $\tauSI$ to be constant, their simulations show shorter periods for larger dust sizes (see their Tables 1 and 2, and Figure \ref{fig:LiYoudin2021} of the present paper). Therefore, the dust growth during the pre-clumping phase might accelerate the clumping. This positive feedback process was also indicated in \citet{Bai2010b} and will operate even if $\rhod/\rhog$ increases only up to a few or $\sim10$ by the moderate clumping (e.g., Z0.6t2 and Z0.3t30 runs in \citet{Li2021}) since dust grains should grow gradually. The efficiency of the feedback will depend on a production rate of larger dust grains due to the coagulation in clumps, which we will address in our future work.

In the present model calculations, we assume equal-mass collisions, where the collision velocity is primarily due to turbulent motion. If we consider unequal-mass collisions, the coagulation timescale may become shorter since dust grains also collide through relative drift motion. We note that collisions in dense clumps are still necessary to avoid fragmentation since collision velocity of unequal-mass dust grains is higher outside the clumps \citep[see][]{Johansen2009b,Bai2010b}. Dense clumps will also help dust grains to overcome the drift barrier since the drift velocity is reduced in dense regions \citep[][]{Nakagawa1986,Bai2010b}. On the other hand, recent studies show that streaming instability is less efficient when one includes a dust size distribution \citep[][]{Krapp2019,Paardekooper2020,McNally2021,Zhu2021,Yang2021}. This may indicate that it takes longer time to achieve the strong clumping. Therefore, coagulation will be even more effective in this case than the clumping, which further motivates us to consider coagulation during the pre-clumping phase. Further quantitative studies are necessary since the impact of the size distribution depends on its slope and shape \citep[][]{Zhu2021,McNally2021}.

In our fiducial case, we assume $T=50\;\mathrm{K}$ and consider icy dust grains (see also the top panels of Figure \ref{fig:tratio_silic} for $T=100\;\mathrm{K}$). Numerical simulations of dust aggregate collisions showed $\vfrag\simeq30-100\;\mathrm{m/s}$ for water ice with a monomer size of $0.1\;\mu\mathrm{m}$ \citep[e.g.,][]{Wada2009,Hasegawa2021}. The fragmentation velocity can be about two times smaller than these values for a monomer size of $\simeq0.2\;\mu\mathrm{m}$, which is in the possible range of monomer sizes observationally indicated by \citet{Tazaki2022}. The assumed $\vfrag$ in this work is lower than these values, meaning that the efficiency of coagulation will be higher than we estimated (see also Figure \ref{fig:dv2_as_r_and_d2g}). Therefore, coagulation should be taken into account when one investigates icy planetesimal formation via streaming instability.

In the case of silicate grains, the coagulation efficiency will be lower (the bottom panels of Figure \ref{fig:tratio_silic}) since usually assumed $\vfrag$ for silicate is lower. Regardless of small $\vfrag$, we show that the collision velocity of the Model 1 is kept below 3 m/s for $\cs\simeq0.8\;\mathrm{km/s}$ in moderately clumping regions with $\rhod/\rhog>10$. This suggests that even silicate dust grains can grow toward the size of $\taus=1$. In the case of the Model 2 (the bottom right panel of Figure \ref{fig:tratio_silic}), silicate dust grains can grow toward $\taus=1$ at the strong clumping phase with $\rhod/\rhog\sim10^3$. This required value of $\rhod/\rhog$ strongly depends on $\vfrag$ (see Figure \ref{fig:dv2_as_r_and_d2g} and Equation (\ref{eq:d2g_for_dv})). The silicate dust will grow more easily during the clumping if $\vfrag$ is $\simeq 6\;\mathrm{m/s}$ for silicate dust with 0.1 $\mu\mathrm{m}$-sized monomers \citep[][]{Wada2009} or higher \citep[][]{Kimura2015,Steinpilz2019}. In our future work, we will address time evolution of dust sizes and collision velocities during the clumping in more detail.

\acknowledgments
We thank Xuening Bai for providing us with their simulation data that we used to verify our collision velocity model and for helpful comments. We also thank the anonymous referee for constructive comments that helped us to improve the manuscript. This work was supported by JSPS KAKENHI Grant Nos. 21K20385 (R.T.T.), 19K03941 (H.T.).  R.T.T. is also supported by RIKEN Special Postdoctoral Researchers Program. 


%




\bibliographystyle{aasjournal}
\bibliography{rttominaga}

%
%
%


\end{document}